\documentclass[10pt,conference]{IEEEtran}
\usepackage[T1]{fontenc}
\usepackage[utf8]{inputenc}
\usepackage[english]{babel}
\usepackage{amsmath,amssymb,amsfonts}
\usepackage{colonequals} %
\usepackage{bm} %
\usepackage{comment} %
\usepackage{microtype}
\usepackage[
    backend=biber,
    style=ieee,
    sortlocale=en_US,
    sortcites=true,
    url=true,
    eprint=true,
    giveninits=false,
    dashed=false,
    minnames=1,
    maxnames=5
]{biblatex}
\usepackage{csquotes} %
\usepackage[inline]{enumitem} %
\usepackage{tabularx} %
\usepackage{booktabs} %
\usepackage[hidelinks]{hyperref} %
\usepackage[noabbrev,capitalize,nameinlink]{cleveref}
\usepackage{xcolor}
\usepackage{graphicx}
\makeatletter
\let\MYcaption\@makecaption
\makeatother
\usepackage{subcaption} %
\makeatletter
\let\@makecaption\MYcaption
\makeatother
\usepackage{wrapfig} %
\usepackage{float}  %
\usepackage{placeins} %
\usepackage{xspace}
\usepackage[ruled,vlined,linesnumbered]{algorithm2e}
\usepackage{fnpct} %
\usepackage{amsthm}
\usepackage{siunitx}
\AtEveryBibitem{%
\clearname{editor}%
\clearfield{series}%
\clearfield{isbn}%
\clearfield{issn}%
\clearfield{day}%
\clearfield{month}%
\clearfield{place}%
\clearlist{location}%
\ifentrytype{software}{%
}{%
\clearfield{url}%
\clearfield{urlyear}%
}%
}
\newtheorem{example}{Example}
\crefname{algocf}{Algorithm}{Algorithms}
\Crefname{algocf}{Algorithm}{Algorithms}
\crefname{constraint}{Constraint}{Constraint}
\Crefname{constraint}{Constraint}{Constraints}
\crefname{condition}{Condition}{Condition}
\Crefname{condition}{Condition}{Conditions}
\newcolumntype{R}{>{\raggedleft\arraybackslash}X}
\newcolumntype{C}{>{\centering\arraybackslash}X}

\definecolor{TUM_blue}{RGB}{0,101,189}
\colorlet{TUM_black}{black}
\colorlet{TUM_white}{white}
\definecolor{TUM_darkblue}{RGB}{0,82,147}
\colorlet{TUM_darkblue100}{TUM_darkblue}
\colorlet{TUM_darkblue80}{TUM_darkblue100!80}
\colorlet{TUM_darkblue50}{TUM_darkblue100!50}
\colorlet{TUM_darkblue20}{TUM_darkblue100!20}
\definecolor{TUM_verydarkblue}{RGB}{0,51,89}
\colorlet{TUM_verydarkblue100}{TUM_verydarkblue}
\colorlet{TUM_verydarkblue80}{TUM_verydarkblue100!80}
\colorlet{TUM_verydarkblue50}{TUM_verydarkblue100!50}
\colorlet{TUM_verydarkblue20}{TUM_verydarkblue100!20}
\colorlet{TUM_darkgrey}{TUM_black!80}
\colorlet{TUM_grey}{TUM_black!50}
\colorlet{TUM_lightgrey}{TUM_black!20}
\definecolor{TUM_beige}{RGB}{218,215,203}
\definecolor{TUM_orange}{RGB}{227,114,34}
\definecolor{TUM_green}{RGB}{162,173,0}
\definecolor{TUM_verylightblue}{RGB}{152,198,234}
\definecolor{TUM_lightblue}{RGB}{100,160,200}
\newcommand{\ie}{i.\,e.\nolinebreak\@\xspace}
\newcommand{\eg}{e.\,g.\nolinebreak\@\xspace}
\newcommand{\N}{\mathbb{N}}%

\renewcommand{\phi}{\varphi}

\title{\textls[-9]{An Abstract Model and Efficient Routing for Logical Entangling Gates on Zoned Neutral Atom Architectures}}
\author{
    \IEEEauthorblockN{Yannick Stade\IEEEauthorrefmark{1}, Ludwig Schmid\IEEEauthorrefmark{1}, Lukas Burgholzer\IEEEauthorrefmark{1}, and Robert Wille\IEEEauthorrefmark{1}\IEEEauthorrefmark{2}}
    \IEEEauthorblockA{\IEEEauthorrefmark{1}%
        Chair for Design Automation,
        Technical University of Munich,
        Munich, Germany
    }
        \IEEEauthorblockA{\IEEEauthorrefmark{2}%
        Software Competence Center Hagenberg GmbH,
        Hagenberg, Austria
    }
    \{yannick.stade, ludwig.s.schmid, lukas.burgholzer, robert.wille\}@tum.de\\
    \href{https://www.cda.cit.tum.de/research/quantum}{www.cda.cit.tum.de/research/quantum}
}
\hypersetup{ %
    pdftitle={An Abstract Model and Efficient Routing for Logical Entangling Gates on Zoned Neutral Atom Architectures},
    pdfsubject={IEEE International Conference on Quantum Computing and Engineering},
    pdfauthor={Yannick Stade, Ludwig Schmid, Lukas Burgholzer, Robert Wille}
}

\begin{document}
\maketitle
\begin{abstract}
    Recent experimental achievements have demonstrated the potential of neutral atom architectures for fault-tolerant quantum computing.
    These architectures feature the dynamic rearrangement of atoms during computation---enabling nearly arbitrary \mbox{two-dimensional} rearrangements.
    Additionally, they employ a zoned layout with dedicated regions for entangling, storage, and readout.
    This architecture requires design automation software that efficiently compiles quantum circuits to this hardware and takes care that atoms are in the right place at the right time.
    In this paper, we initiate this line of work by providing,
    \begin{enumerate*}[label=(\arabic*)]
        \item an abstract model of the novel architecture and,
        \item an efficient solution to the routing problem of entangling gates.
    \end{enumerate*}
    By this, we aim to maximize the parallelism of entangling gates and minimize the overhead caused by the routing of atoms between zones.
    In addition to that, we keep the realm of
    fault-tolerant quantum computing in mind and consider logical qubit arrays, each of which encodes one logical qubit.
    We implemented the proposed idea as a tool called NALAC and demonstrated its effectiveness and efficiency by showing that it can significantly reduce the routing overhead of logical entangling gates compared to the naive approach.
    As part of the \emph{Munich Quantum Toolkit}~(MQT), NALAC is publicly available as open-source at \url{https://github.com/cda-tum/mqt-qmap}.
\end{abstract}
\section{Introduction}\label{sec:introduction}
Over the past decade, significant efforts have been devoted to developing first, intermediate-scale quantum computing demonstrators and mitigating physical errors~\cite{preskillQuantumComputingNISQ2018}---advancing both hardware and software development.
Despite ongoing reductions in hardware errors, achieving the requisite error rates, on the order of $10^{-10}$~\cite{gidneyHowFactor20482021}, necessary for executing large-scale meaningful algorithms such as integer factorization will likely require the implementation of sophisticated \emph{Quantum Error Correction}~(QEC) protocols~\cite{preskillQuantumComputingNISQ2018}.
This transition to \emph{Fault-Tolerant Quantum Computing}~(FTQC) necessitates advanced experimental setups that meet the novel requirements and constraints imposed by FTQC, alongside appropriate software support, to maximize the utilization of available hardware capabilities.

Experimental progress has demonstrated elementary FTQC operations on various hardware platforms, including superconducting chips~\cite{acharyaSuppressingQuantumErrors2023,barendsSuperconductingQuantumCircuits2014,takitaExperimentalDemonstrationFaultTolerant2017,zhaoRealizationErrorCorrectingSurface2022}, trapped ions~\cite{ryan-andersonImplementingFaulttolerantEntangling2022,eganFaulttolerantControlErrorcorrected2021,postlerDemonstrationFaulttolerantUniversal2022a,pogorelovExperimentalFaulttolerantCode2024}, and neutral atoms~\cite{bluvsteinLogicalQuantumProcessor2023}.
However, these experiments often involve small, hand-constructed examples.
In order to scale up to sizes relevant for practical use, sophisticated design automation methods and software are essential~\cite{burgholzerDesignAutomationTools2023}.

Neutral atoms have emerged as a promising platform for universal quantum computing~\cite{bluvsteinLogicalQuantumProcessor2023,henrietQuantumComputingNeutral2020,saffmanQuantumInformationRydberg2010,saffmanQuantumComputingNeutral2019,grahamMultiqubitEntanglementAlgorithms2022}, offering
long coherence times, arbitrary connectivity through dynamic atom rearrangement, and highly parallel gate execution between sets of atoms.
Recent experimental breakthroughs
were achieved using a novel zoned architecture, where different functionalities (storage, entangling, measurement) are performed in designated spatially separated zones, with shuttling facilitating the transfer of atoms between those zones~\cite{bluvsteinLogicalQuantumProcessor2023}.
Despite the rapid development of neutral atom-specific compilation software~\cite{schmidComputationalCapabilitiesCompiler2023} and routing of atom movements~\cite{tanCompilingQuantumCircuits2023,wangFPQACCompilationFramework2023,wangQPilotFieldProgrammable2023,nottinghamDecomposingRoutingQuantum2023,schmidHybridCircuitMapping2023}, the absence of an appropriate abstract model and the lack of dedicated routing solutions for logical qubit arrays hinders the efficient utilization of this novel architecture and its capabilities.

In this work, we pioneer this line of work two-fold by
\begin{enumerate}
    \item presenting an abstract model of the zoned architecture introduced in~\cite{bluvsteinLogicalQuantumProcessor2023}, including its capabilities, constraints, as well as the resulting problem of routing logical entangling gates, and
    \item proposing an efficient approach to tackle this routing problem by reducing routing overhead and increasing gate parallelism.
\end{enumerate}

To this end, we utilize graph theoretical techniques to derive a highly efficient solution for the routing problem.
The core of the proposed approach is a specific coloring of the interaction graph representing the entangling gates in the quantum circuit.
Thereby, compared to a naive solution, we schedule multiple entangling gates in parallel and ensure that multiple sets of entangling gates are applied without intermediate time-expensive shuttling of atoms between zones.
We implemented the approach in the tool NALAC (\emph{Neutral Atom Logical Array Compiler}) and benchmarked it on different quantum circuits, demonstrating its efficiency and effectiveness.

The model of the neutral atom architecture and the proposed compiler NALAC pave the road for further compiler development on similar architectures and introduce the first software tool providing an automated solution to the routing problem of transversal logical gates on zoned neutral atom architectures.
The tool NALAC is integrated into the \emph{Munich Quantum Toolkit}~(MQT,~\cite{mqt}) that comprises simulators~\cite{burgholzerSimulationPathsQuantumCircuit2023}, validation tools~\cite{pehamEquivalenceCheckingParameterized2023}, and tools to automatically select optimization passes~\cite{quetschlichMQTPredictorAutomaticDeviceSelection2024} among many others.
The full code of the proposed approach is publicly available at \mbox{\url{https://github.com/cda-tum/mqt-qmap}}.

The remainder of this paper is structured as follows: \Cref{sec:preliminaries} provides background on FTQC and a brief introduction to neutral atoms.
In \cref{sec:zoned-neutral-atom-model}, we introduce the model of the zoned architecture together with its capabilities and constraints.
\Cref{sec:logical-array-routing} discusses the architecture-specific routing problem and illustrates the proposed solution compared to a naive approach.
Technical implementation details are provided in Section~\ref{sec:implementation-details}.
\Cref{sec:evaluation} evaluates the proposed approach on a set of benchmark circuits, demonstrating reduced routing overhead.
Finally, \cref{sec:conclusions} concludes the paper.

\section{Preliminaries}\label{sec:preliminaries}

In order to keep the remainder of this work \mbox{self-contained}, this section briefly describes the concept of logical quantum computing, which is essential when employing \emph{Fault-Tolerant Quantum Computing}~(FTQC).
It then delves into the physical background of quantum computing using neutral atoms.

\subsection{Fault-Tolerant Quantum Computing}\label{sec:logical-qc}

The overarching principle of FTQC is the idea of distributing the information of a single \emph{logical} qubit~$q^{L}$ across multiple \emph{physical} qubits $q_{1},\dots,q_{n}$.
The resulting redundancy enables the detection of individual errors at the physical level and appropriate corrections~\mbox{\cite{shorFaulttolerantQuantumComputation1996,gottesmanStabilizerCodesQuantum1997}}.
We henceforth refer to a \emph{(logical) qubit array} as the set of physical qubits used to encode a single logical qubit.
This is illustrated by means of the popular Stean code as the simplest example of the widely used color code~\cite{bombinGaugeColorCodes2015}.

\begin{example}[Logical Qubit Arrays]\label{exp:logical-arrays}
    The Stean code requires $n=7$ physical qubits $q_{1},\dots,q_{7}$ to encode a single logical qubit $q^{L}$.
    Below, we illustrate the physical qubits arranged, \eg, in a $2\times4$ array.
    Throughout this paper, we will generally use squares to represent logical qubits and circles for physical qubits.
    \begin{figure}[H]
        \centering
        \vspace{-4pt}
        \includegraphics[width=\linewidth]{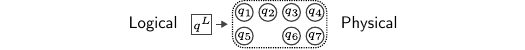}\\
        \vspace{-4pt}
    \end{figure}
\end{example}

The execution of operations on these encoded logical qubits (indicated by superscripts, \eg, $\mathrm{CZ}^{L}$) depends on the specific QEC code and is generally highly complex---often requiring procedures such as braiding or magic state injection~\cite{fowlerSurfaceCodesPractical2012}.
This study focuses on the special case of \emph{transversal} gates, where the logical operation is executed by performing the same operation on each physical qubit individually.
This prevents the possible spreading of errors, making transversal gates inherently fault-tolerant.
In particular, we focus on codes that provide transversal entangling gates (such as CX, CZ, $\dots$), encompassing surface codes~\cite{fowlerSurfaceCodesPractical2012}, color codes~\cite{bombinGaugeColorCodes2015}, and, in general, the family of \emph{Calderbank-Shor-Stean}~(CSS) codes~\cite{shorFaulttolerantQuantumComputation1996}.
\begin{example}[Transversal CZ gate]
    To implement a logical $\mathrm{CZ}^{L}$ gate that is transversal for the respective code, a physical CZ gate must be applied between all pairs of the respective physical qubits within the two logical arrays:
    \begin{figure}[H]
        \centering
        \vspace{-4pt}
        \includegraphics[width=\linewidth]{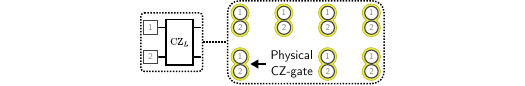}\\
        \vspace{-4pt}
    \end{figure}
\end{example}

Importantly, transversal two-qubit gates alone are insufficient for achieving universal fault-tolerant computation~\cite{eastinRestrictionsTransversalEncoded2009}.
Nevertheless, due to their inherent fault tolerance, they are a fundamental building block, essential for large-scale fault-tolerant computing. %
Therefore, correct handling of transversal gates constitutes the first important step toward a full compilation stack for FTQC.

\subsection{Neutral Atom Quantum Computing}\label{sec:neutal-atom-quantum}
Quantum computing based on \emph{neutral atoms}~\cite{saffmanQuantumInformationRydberg2010,saffmanQuantumComputingNeutral2019,henrietQuantumComputingNeutral2020,schmidComputationalCapabilitiesCompiler2023,acharyaSuppressingQuantumErrors2023,grahamMultiqubitEntanglementAlgorithms2022} relies on qubits encoded into long-lived atomic states of alkali or alkaline-earth(-like) atoms such as Rb, Sr, or Yb.
Atoms are confined within optical dipole traps generated by optical lattices or optical tweezers and, subsequently, laser-cooled to their motional ground state, facilitating the efficient trapping of thousands of atoms~\cite{manetschTweezerArray61002024,barredoAtombyatomAssemblerDefectfree2016,pauseSuperchargedTwodimensionalTweezer2024}.
Recent experiments demonstrate the continuous reloading of atoms~\cite{gygerContinuousOperationLargescale2024,norciaIterativeAssembly1712024} to compensate for potential losses. %

\mbox{Single-qubit} gates are implemented by inducing state transitions using either global or focused laser beams, addressing the entire register or individual atoms~\cite{wangIndividualatomControlArray2023,bluvsteinLogicalQuantumProcessor2023,grahamMultiqubitEntanglementAlgorithms2022,levineDispersiveOpticalSystems2022}, respectively.
Entangling gates exploit high-lying Rydberg states and their long-range dipole-dipole interactions among proximal atoms via Rydberg blockade~\cite{mullerMesoscopicRydbergGate2009,isenhowerDemonstrationNeutralAtom2010,saffmanQuantumInformationRydberg2010}.
Notably, parallel entanglement of numerous qubits is achieved by globally illuminating the qubit register, inducing two-qubit gates between pairs of qubits within the interaction range of the Rydberg blockade~\cite{levineParallelImplementationHighFidelity2019,everedHighfidelityParallelEntangling2023}.
Measurements are realized using fluorescence imaging or other non-destructive measurement techniques to read out qubit states in the computational basis~\cite{grahamMidcircuitMeasurementsNeutral2023,norciaMidcircuitQubitMeasurement2023,deistMidCircuitCavityMeasurement2022}.

\begin{example}[Native Gates]\label{exp:native-gates}
    \begin{figure}[tb]
        \centering
        \vspace{-3pt}
        \includegraphics[width=\linewidth,trim=0 0 0 0]{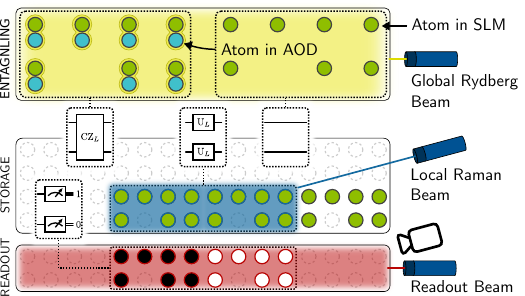}\\
        \vspace{-1pt}
        \caption{
            Zoned architecture considered within this work based on the experimental setup from~\cite{bluvsteinLogicalQuantumProcessor2023}.
        }
        \label{fig:na-operations}
    \end{figure}

    \Cref{fig:na-operations} illustrates the implementation of single- and two-qubit gates utilizing Raman (blue, middle) and Rydberg (yellow, top) lasers, respectively.
    Regarding the Rydberg beam, it is worth noting that qubits nearby execute a CZ gate (top left), while isolated qubits undergo an identity operation (top right).
\end{example}

\begin{figure*}[tb]
    \centering
    \begin{subfigure}{175pt}
        \centering
        \includegraphics[width=175pt]{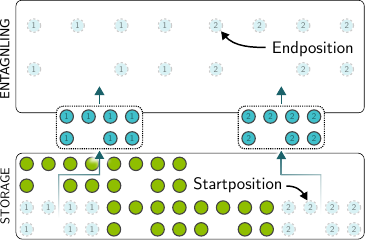}
        \caption{\(t=1\)}
        \label{fig:shuttle set 1}
    \end{subfigure}
    \hspace{-4pt}
    \begin{subfigure}{168pt}
        \centering
        \includegraphics[width=168pt]{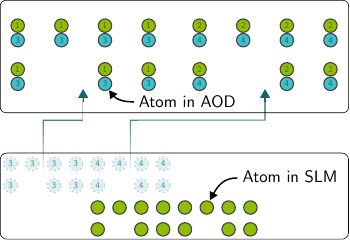}
        \caption{\(t=2\)}
        \label{fig:shuttle set 2}
    \end{subfigure}
    \hspace{-4pt}
    \begin{subfigure}{168pt}
        \centering
        \includegraphics[width=168pt]{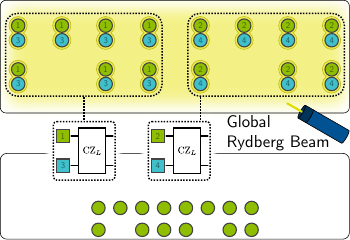}
        \caption{\(t=3\)}
        \label{fig:rydberg beam}
    \end{subfigure}
    \caption{
        One execution cycle up to the execution of the CZ gate as described in \cref{exp:one-execution-cycle}.
    }
    \label{fig:architecture}
\end{figure*}

Novel experimental techniques~\cite{bluvsteinQuantumProcessorBased2022,bluvsteinLogicalQuantumProcessor2023} have demonstrated the ability to \emph{shuttle} atoms during computation, \ie, to change the position of trapped atoms, allowing almost arbitrary two-dimensional rearrangements (this will be covered in more detail in \cref{sec:shuttling constraints}).
This is achieved by introducing two possible types of optical traps between which atoms can switch: First, static \emph{Spatial Light Modulator}~(SLM) traps afford placement in nearly arbitrary yet fixed configurations; secondly, a dynamically adjustable 2D optical lattice, employing two \emph{Acousto-Optic Deflectors}~(AODs), arranged along the x- and y-axes, enables the parallel rearrangement of multiple qubits.

Overall, neutral atom architectures offer great parallelism for gate execution, which is favorable for transversal gates.
The laser beams can illuminate entire qubit registers, inducing parallel operations on all illuminated qubits.
Thus, a transversal operation on a logical qubit can be applied to all corresponding atoms in parallel.

\section{Abstract Model\\of Zoned Neutral Atom Architectures}\label{sec:zoned-neutral-atom-model}
So far, previous work has considered a monolithic architecture where all operations are performed in the same region~\cite{bluvsteinQuantumProcessorBased2022}.
In the recent experimental milestone~\cite{bluvsteinLogicalQuantumProcessor2023},
different operations are conducted in spatially separated zones.
Quantum circuits are executed by alternatingly shuttling atoms between zones and illuminating them with the corresponding laser beams.
In order to scale to large quantum circuits, the compilation process cannot be performed manually anymore, and design automation software is needed.
However, this is still lacking as no abstract model for zoned architecture that captures its capabilities and constraints exists.
To this end, we first review the zoned architecture based on the experimental setup from~\cite{bluvsteinLogicalQuantumProcessor2023} and the execution steps.
Based on that, we then formalize the resulting shuttling constraints and requirements for executing logical, transversal gates---eventually resulting in the required model.

\subsection{Zoned Structure}\label{sec:zoned-neutral-atom}
In general, zoned architectures spatially separate certain operations, allowing favorable hardware setups and optimization.
As depicted in \cref{fig:na-operations}, the architecture considered in this work consists of three separate zones, each specifically designed for a particular purpose:
\begin{enumerate}[label=\alph*)]
    \item \textbf{Entangling zone:}
          This zone is dedicated to the execution of entangling operations, \ie, CZ gates in the case of~\cite{bluvsteinLogicalQuantumProcessor2023}.
          Using a global Rydberg beam applied to the whole zone, the atoms are excited to the Rydberg state.
          This applies parallel CZ entangling gates between all qubit pairs within the range of the Rydberg blockade of each other.
          Qubits not supposed to interact are placed sufficiently far from each other so that the Rydberg interaction can be neglected.
          A single qubit without another qubit nearby is still excited to the Rydberg state, but without an interaction partner, it returns to its original state, effectively performing an identity operation.
          However, the Rydberg decay might possibly introduce an error similar to an actual CZ gate.
    \item \textbf{Storage zone:}
          This zone allows for densely packaging qubits and application of \mbox{single-qubit} gates while maintaining high coherence times.
          In this zone, the qubits are shielded from beams that could potentially disturb the quantum state (Rydberg or measurement).
          A large number of static SLM traps provide the necessary storing opportunities.
          Single-qubit rotations are realized using global and/or individually addressable laser beams.
    \item \textbf{Readout zone:}
          In this zone, the state of a set of qubits can be read out without disturbing the quantum state of other qubits.
          In particular, this enables mid-circuit measurements, which are essential for full-stack FTQC.
          Applying a readout beam measures all qubits within the readout zone simultaneously.
\end{enumerate}

\begin{figure*}[tb]
    \centering
    \begin{subfigure}{\linewidth}
        \centering
        \includegraphics[width=\linewidth]{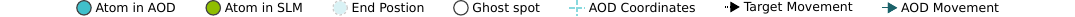}
    \end{subfigure}
    \begin{subfigure}[t]{.49\linewidth}
        \centering
        \includegraphics[width=.95\linewidth]{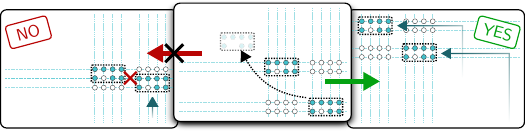}
        \caption{
            Row-Column Non-Crossing Constraint%
        }
        \label{fig:shuttling constraint non cross}
    \end{subfigure}
    \begin{subfigure}[t]{.49\linewidth}
        \centering
        \includegraphics[width=.95\linewidth]{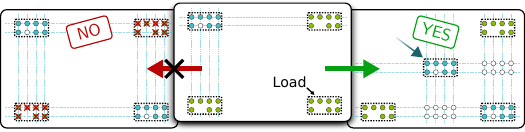}
        \caption{
            Row-Column Preservation Constraint%
        }
        \label{fig:shuttling constraint preservation}
    \end{subfigure}
    \begin{subfigure}[t]{.49\linewidth}
        \centering
        \includegraphics[width=.95\linewidth]{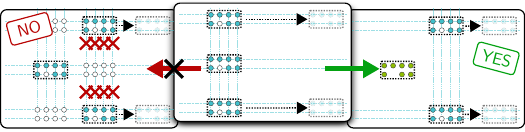}
        \caption{
            Avoiding Ghost-Spots Constraint%
        }
        \label{fig:shuttling constraint ghost}
    \end{subfigure}
    \begin{subfigure}[t]{.49\linewidth}
        \centering
        \includegraphics[width=.95\linewidth]{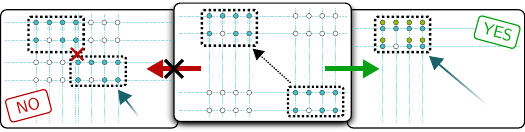}
        \caption{
            Array Alignment Constraint%
        }
        \label{fig:shuttling constraint array}
    \end{subfigure}
    \caption{
        Illustration of the four constraints on shuttling operations.
    }
    \label{fig:aod-constraints}
\end{figure*}

\subsection{Quantum Circuit Execution Steps}\label{sec:quant-circ-exec}
While zoned architectures can yield improved hardware characteristics by shielding stored and operated qubits from each other, shuttling is required to move the respective qubits into the desired zones.
Initially, the qubits are located in the storage zone, providing, \eg, the best coherence times.
Given a sequence of transversal gates, single-qubit operations can be executed immediately without additional shuttling.
However, as described above, CZ gates can only be performed in the entangling zone.
Therefore, the task at hand is to shuttle qubits from the storage to the entangling zone, perform the CZ gate, and shuttle them back to the storage zone.
In particular, for logical qubits, this requires the movement of all physical qubits corresponding to the logical qubit.
One shuttle operation that moves (multiple) qubits from one SLM trap to another consists of the following steps:
\begin{enumerate}
    \item\label{itm:shuttle first} \textbf{Loading:} Qubits are loaded from static SLM traps into adjustable AOD traps.
    \item \textbf{Moving:} The loaded qubits can be moved in the inter-atomic space without disturbing other atoms.
    \item\label{itm:shuttle last} \textbf{Storing:} Finally, the qubits are transferred from AOD traps back to SLM traps.
\end{enumerate}
If not required, the last step can also be skipped, and qubits can remain in adjustable AOD traps.
After the required qubits are shuttled to the entangling zone and arranged such that pairs of atoms that are supposed to interact are located in proximity, the CZ gate is applied by switching on the Rydberg beam.

\begin{example}[One Execution Cylce]\label{exp:one-execution-cycle}
    \Cref{fig:architecture} sketches the steps necessary to perform two parallel CZ gates between the qubit-pairs \((1,3)\) and \((2,4)\): %
    \begin{enumerate}[label=\(\mathbf{t=\arabic*}\),leftmargin=*]
        \item Qubit arrays 1 and 2 are transferred from the storage to the entangling zone in parallel.
              This includes loading, moving, and storing %
              (the necessity of storing is discussed in \cref{sec:shuttling constraints}).
        \item The next set of qubit arrays corresponding to qubits 3 and 4 are moved to the entangling zone in parallel.
        \item The global Rydberg beam (yellow) performs entangling CZ gates between all qubit pairs of the arrays \((1,3)\) and \((2,4)\), respectively.
              Afterward, the previous steps can be applied in reverse order to move the atoms back to the storage zone.
    \end{enumerate}
\end{example}

Similarly, that process can also transport qubits between the storage and readout zones.
By consistently applying steps (\ref{itm:shuttle first})--(\ref{itm:shuttle last}), it becomes possible to execute arbitrary sequences of transversal gates through array shuttling between zones and activation of laser beams.

\subsection{Shuttling Constraints}\label{sec:shuttling constraints}

However, an AOD cannot load and move the qubits arbitrarily; the shuttling of qubits must follow certain constraints~\cite{bluvsteinQuantumProcessorBased2022}.
As discussed in \cref{sec:neutal-atom-quantum}, atoms can be confined in static SLM traps or dynamically adjustable AODs.
Each AOD generates a series of laser beams (rows and columns) along the x- or y-axis at precise coordinates, capable of individual activation and deactivation.
The intersections of these orthogonal beams form a grid that defines trap coordinates within the 2D plane, allowing for the manipulation of trap positions by modulating beam coordinates and consequently moving the trapped atoms.
This method facilitates the trapping and simultaneous movement of a larger quantity of atoms, provided they adhere to the following four constraints, which later will be illustrated in \cref{exp:shuttling constraints}:
\begin{enumerate}[label=\alph*), ref=(\alph*)]
    \item\label[constraint]{itm:aod constraint non cross} \textbf{Row-Column Non-Crossing Constraint}: AOD rows (columns) must maintain a minimum separation during movements and must not intersect with each other.
          In particular, the order among rows (columns) must be preserved during movements.
    \item\label[constraint]{itm:aod constraint preservation} \textbf{Row-Column Preservation Constraint}: If atoms are in the same row (column), they remain in the same row (column) throughout a movement.
    \item\label[constraint]{itm:aod constraint ghost} \textbf{Avoiding Ghost-Spots Constraint}: Each intersection of AOD rows and columns generates a potential trap, resulting in a grid of traps.
          Some of those traps might be unintended, so-called \emph{ghost spots} and can unintentionally trap or disturb other atoms. %
    \item\label[constraint]{itm:aod constraint array} \textbf{Array Alignment Constraint}: To perform a transversal logical CZ gate between two arrays, they must overlap to bring the respective qubit pairs together.
          With both arrays loaded in the same AOD, this would violate \cref{itm:aod constraint non cross}.
          Therefore, one of the two arrays has to be placed back in SLM traps, which are not subject to the AOD constraints.
\end{enumerate}
\begin{example}[Shuttling Constraints]\label{exp:shuttling constraints}
    \Cref{fig:shuttling constraint non cross} depicts the case where the lower right array is supposed to move to the upper left corner.
    In order to make this possible, the middle atom array cannot remain at its position and has to be moved to the upper left as well to avoid AOD crossing.
    Alternatively, the middle array could be (temporarily) stored in SLM traps.

    \Cref{fig:shuttling constraint preservation} illustrates the case where the middle array is supposed to stay at its position while the other two should move right.
    Before moving the other two arrays, the middle array must be stored in SLM traps.

    Loading the lower right array shown in \cref{fig:shuttling constraint ghost} cannot be done directly without simultaneously loading the other two arrays.
    An additional offset movement is required to move the ghost spots into the inter-atomic space to load the array in a second step.

    \Cref{fig:shuttling constraint array} illustrates that at least one of the two arrays of a logical entangling gate has to be placed in SLM traps for the arrays to overlap.
    Otherwise, the AOD crossing constraint will be violated.
\end{example}

With the zoned structure and the constraints in place, the zoned neutral atom architecture model is complete.
Now, the model can be used to formulate the routing problem on this architecture.
For the special case of FTQC, we describe in the next section how the physical architecture can be abstracted into a logical model.

\subsection{Towards Fault-Tolerant Quantum Computing}\label{sec:ftqc}

We want to pave the way to FTQC, supporting larger atom arrays that encode a logical qubit.
The logical arrays must be shuttled between the zones to execute the different operations and establish the necessary connectivity %
for entangling gates.
For the considered sequence of transversal gates, it suffices to compute the movements on the logical level, \ie, for the logical qubits representing an array of physical qubits.
This is possible since all physical qubits of one logical array can be shuttled simultaneously.

\begin{figure}[tb]
    \centering
    \includegraphics[width=\linewidth]{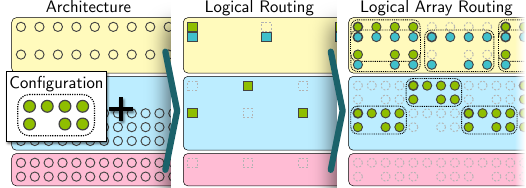}
    \caption{
        The physical architecture and the configuration (left), \ie, array size for the code, serve as the input.
        A logical architecture is derived from this and used to find a logical routing (middle).
        This logical routing is then translated into a logical array routing (right).
    }
    \label{fig:ftqc}
\end{figure}

Using a configuration that specifies the array size for the code, the physical architecture can be abstracted to a logical architecture.
The physical architecture is parqueted with non-overlapping arrays of qubits, and only the left upper qubit of each array is taken as a reference for the logical qubit.
After all movements necessary for the circuit execution have been computed, the logical qubits are replaced by their corresponding array of physical qubits.
This process is illustrated in \cref{fig:ftqc}.

\section{Routing of Logical Entangling Gates\\for Zoned Architectures}\label{sec:logical-array-routing}

Shuttling qubits into proximity for entangling gates while simultaneously minimizing overhead through loading, storing, and shuttling constitutes a significant focus of ongoing research in neutral atom compilation.
Different approaches have been proposed for different, non-zoned architectures~\cite{brandhoferOptimalMappingNearTerm2021,tanCompilingQuantumCircuits2023,nottinghamDecomposingRoutingQuantum2023,wangQPilotFieldProgrammable2023,wangFPQACCompilationFramework2023,schmidHybridCircuitMapping2023}.
Other zone-based solutions~\cite{schoenbergerShuttlingScalableTrappedIon2024,kreppelQuantumCircuitCompiler2023,schoenbergerUsingBooleanSatisfiability2024,webberEfficientQubitRouting2020} are targeted towards trapped ion hardware which has fundamentally different shuttling constraints~\cite{hensingerQuantumComputerBased2021,muraliArchitectingNoisyIntermediateScale2020}.
For neutral atoms, however, we first define the considered problem, present the straightforward, naive solution, and then introduce the proposed solution to the routing problem.

\subsection{Considered Problem}\label{sec:routing-problem}
The model proposed in the previous section can now be used to formulate the problem of \emph{routing of logical, transversal entangling gates} that compilers have to solve for zoned neutral atom architectures to generate a sequence of execution steps.
More precisely, given a sequence of entangling \mbox{(2-qubit)}, transveral gates, the goal is to move the qubit arrays to the required zone for the subsequent operation with minimal \emph{routing time overhead}---that is, the additional time required for the qubit movement---while adhering to the constraints outlined in \cref{sec:shuttling constraints}.
According to \cref{sec:ftqc}, it suffices to solve the routing problem on the logical level; hence, whenever we refer to qubits below, we mean logical qubits.

\subsection{Naive Solution}\label{subsec:naive solution}

A simple solution to generate an executable sequence of instructions avoiding AOD crossings is to execute one entangling operation at a time and shuttle the qubits to the entangling zone one by one.
Idling qubits are kept in the storage zone to shield them from the interaction with the Rydberg beam.
For each entangling operation on a pair of qubits, the naive solution moves them to the entangling zone, executes the entangling operation, and moves them back to the storage zone.
In order to satisfy the AOD constraints, one of the qubits is moved first and stored in an SLM trap before the other one is loaded.
The latter is moved close to the first to be in the interaction radius.

This naive solution serializes the execution of the entangling operations. %
Consequently, it does not exploit the potential parallelism of the neutral atom architecture.

\subsection{Proposed Solution}\label{sec:proposed solution}
Motivated by the shortcomings of the naive solution, this work introduces an alternative approach that tries to maximize the number of parallel entangling operations.
This section outlines the general idea of the proposed solution while the details of its realization follow in \cref{sec:implementation-details}.
We improve the naive solution by addressing two primary objectives:
\begin{enumerate}
    \item \textbf{Parallel gate execution:} Execute as many independent entangling operations as possible per Rydberg beam to achieve high gate parallelism.
    \item \textbf{Parallel AOD shuttling:} Load, store, or move as many qubits in parallel to minimize routing time overhead while fulfilling all AOD constraints from \cref{sec:shuttling constraints}.
\end{enumerate}

We accomplish this by splitting the execution of the entangling gates into multiple \emph{runs}.
A run starts with shuttling a set of qubits into the entangling zone and ends with shuttling them back to the storage zone.
Recall that having one qubit of each interacting pair in the SLM trap is necessary but also sufficient to comply with the array alignment constraint~(\cref{itm:aod constraint array}).
This fact can be exploited to execute multiple entangling operations in one run without additional load or store operations.
Therefore, a run is divided into \emph{steps}, such that one step executes a set of independent entangling gates, where independent means that no two entangling gates share any qubits.
The qubits held in the AOD move between the steps to their next interaction partner or resting position.
This way, a two-fold increase in parallelism is achieved: First, by executing multiple entangling operations in parallel within one step, and second, by shuttling additional qubits in parallel into the entangling zone required for the steps in one run.

\begin{figure}[b]
    \centering
    \includegraphics[width=\linewidth]{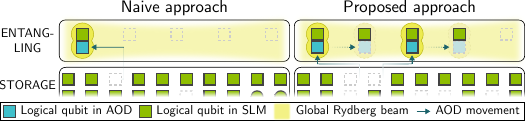}
    \caption{
        Comparison of sequential entangling operations (left) with proposed parallel execution and efficient logical qubit movement (right). %
    }
    \label{fig:naive vs proposed approach}
\end{figure}

\begin{figure*}[t]
    \centering
    \includegraphics[width=\linewidth]{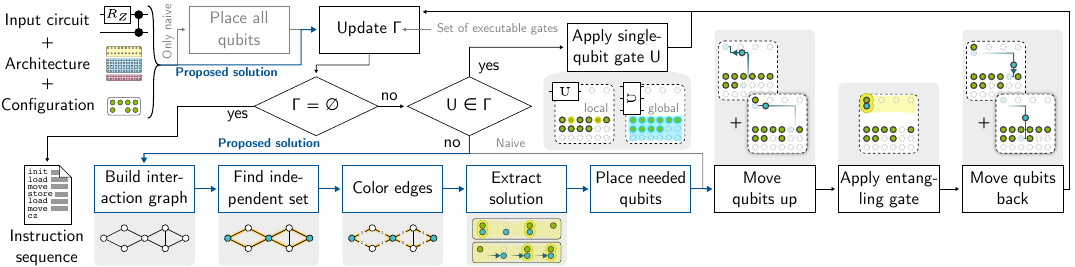}
    \caption{
        This flowchart illustrates the combination of the naive and the proposed solution.
        Where both solutions differ, the naive one is drawn in light grey and the proposed one in dark blue; common parts in black.
        All gates (CZ, single-qubit) are assumed to be transversal.
    }
    \label{fig:general idea}
\end{figure*}

\begin{example}[Naive vs. Proposed Solution]\label{exp:naive vs proposed approach}
    \Cref{fig:naive vs proposed approach} (left) illustrates the naive solution, which executes only one entangling operation at a time.
    In contrast, \cref{fig:naive vs proposed approach} (right) illustrates the proposed solution, which executes two entangling operations in parallel.
    The proposed solution has also placed further qubits into SLM traps in the entangling zone to execute more entangling operations without additional load or store operations.
    For the next step, the qubits move one position to the right to execute two more entangling operations.
    Afterward, all qubits will be moved back to the storage zone.
\end{example}

Independent operations must be identified to realize this parallel execution of entangling gates.
To this end, an interaction graph with qubits as nodes and edges between qubits that should interact is used and provides the core of the proposed approach.
Based on this graph, we compute a maximal independent set that decides which qubits are placed in SLM and AOD traps.
In order to obtain a feasible run regarding the AOD constraints, the edges of the interaction graph are colored such that the color determines the step in which the entangling operation represented by the edge is executed.
Generally, the proposed approach proceeds as illustrated in \cref{fig:general idea}.
The following section will explain each step in detail.

\section{Implementation Details}\label{sec:implementation-details}
This section provides technical details of the proposed approach.
We detail the construction of the interaction graph and how we obtain a schedule for the entangling operations from it by coloring its edges.
Finally, we derive the exact positions of the qubits for each step.

\subsection{Interaction Graph}\label{sec:interaction graph}
The interaction graph is an undirected graph, where the nodes represent qubits that are involved in an entangling operation, and an edge connects two nodes if the corresponding qubits share one entangling operation that is executable at this moment.
An entangling operation is executable if all operations in front are either already executed or commute with the entangling operation.

\begin{figure}[b]
    \centering
    \includegraphics[width=\linewidth]{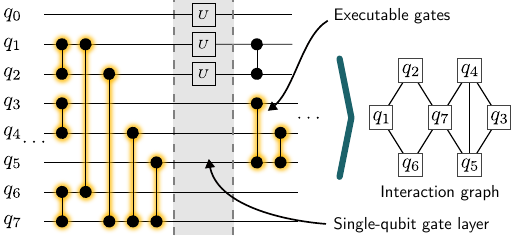}
    \caption{
        Part of a quantum circuit with its currently executable gates highlighted in orange and its corresponding interaction graph on the right.
    }
    \label{fig:interaction graph}
\end{figure}
\begin{example}[Interaction Graph]\label{exp:interaction graph}
    On the left, \cref{fig:interaction graph} shows part of a quantum circuit acting on eight (logical) qubits.
    All hidden gates to the left have already been executed, and all currently executable gates are highlighted in orange.
    Note that two of the three CZ gates to the right of the layer of single-qubit gates are executable because no single-qubit gate acts on their qubits.
    On the right, the figure shows the resulting circuit's interaction graph.
\end{example}

\subsection{Independent Set}\label{subsec:independent set}
For every entangling gate, \ie, edge, that we execute, one of the adjacent qubits needs to be in an SLM trap and the other in an AOD trap.
While trying to maximize the number of entangling gates, this leads to a max-cut problem.
In general, a cut of a graph is a partition of the nodes into two sets.
Accordingly, a max-cut is a cut that maximizes the number of edges between the two sets.

In our case, one set of nodes in the resulting cut are the qubits held in SLM traps, and the other set are qubits in AOD traps.
The problem of finding a max-cut is NP-hard, and we use a maximal independent set instead to find a sufficiently good solution.

An independent set of nodes in a graph is a set of nodes that are not adjacent to each other.
An independent set is maximal if it cannot be extended by adding another node without violating the independence property.
The independent set then constitutes one set of the cut, and the remaining nodes the other set.
Such a maximal independent set can be calculated by iterating over the nodes in the interaction graph and adding one node at a time to the set if it is not adjacent to any of the already selected nodes.
To maximize the number of edges in the resulting cut, we order the nodes by their degree in descending order before iterating over them.
All nodes ending up in the independent set correspond to the qubits kept in the AOD.
The edges covered by the independent set, \ie, those that are adjacent to one node in the independent set, are the ones that can be executed in one run.

\begin{example}[Independent Set]\label{exp:independent set}
    \begin{figure}[tb]
        \centering
        \includegraphics[width=\linewidth]{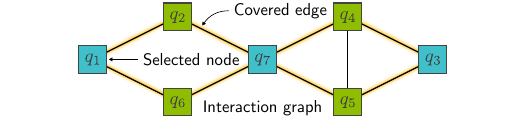}
        \caption{
            The interaction graph from \cref{exp:interaction graph} with the nodes of the independent set colored in blue.
            All edges covered by this independent set are highlighted in orange.
        }
        \label{fig:independent set}
    \end{figure}
    Consider again the interaction graph discussed in \cref{exp:interaction graph}.
    Here, the node with the highest degree is the middle one and is selected first.
    The two nodes with degree three are not selected since they are adjacent to the previously selected one.
    Finally, the algorithm selects the two nodes with degree two that are not adjacent to the selected node.
    The resulting independent set is shown in \cref{fig:independent set}, where the selected nodes are colored in blue.
    Those are the ones that will be held in the AOD, whereas the green ones will be placed in SLM traps.
\end{example}

\subsection{Color Edges}\label{subsec:color edges}
To determine the order in which the entangling operations are executed, we color the covered edges of the interaction graph.
A color is a non-negative integer that is assigned to an edge and corresponds to the time step in which the entangling operation is executed.
Consequently, the number of required colors corresponds to the number of time steps needed to execute all entangling operations.
Two adjacent edges, \ie, edges that share a node, cannot have the same color.
However, another constraint is necessary for a valid coloring to prevent AOD crossings.

\begin{example}[Order Preservation of AOD Qubits]\label{exp:non-crossing constraint}
    Consider four qubits where each pair of \((q_1,q_2)\), \((q_2,q_3)\), \((q_3,q_4)\), and \((q_4,q_1)\) should undergo a CZ operation.
    The upper time sequence in the figure below is impossible because the two AOD qubits would need to cross each other, conflicting with the AOD \cref{itm:aod constraint non cross} from \cref{sec:shuttling constraints}.
    As shown in the lower time sequence, the AOD qubits must maintain their order to prevent crossings, which makes three steps necessary to execute those four CZ operations.
    \begin{figure}[H]
        \centering
        \vspace{-4pt}
        \includegraphics[width=\linewidth]{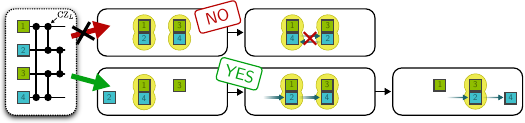}\\
        \vspace{-4pt}
    \end{figure}
\end{example}
In terms of the coloring of the interaction graph's edges, this can be achieved by adding the following condition:
\begin{enumerate}[label=(C)]
    \item\label[condition]{non-crossing constraint} On all paths of length two between two blue nodes, the inequality relation between the assigned colors must point towards the same (blue) end node.
\end{enumerate}
\begin{example}[Coloring Constraint]\label{exp:coloring constraint}
    The two execution sequences above correspond to the two possible colorings below.
    The left coloring violates \cref{non-crossing constraint} whereas the right one satisfies it.
    \begin{figure}[H]
        \centering
        \vspace{-4pt}
        \includegraphics[width=\linewidth]{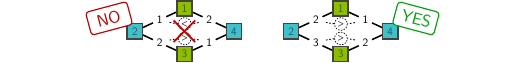}\\
        \vspace{-4pt}
    \end{figure}
\end{example}

To satisfy this additional constraint while keeping the total number of colors (\ie, time steps) low, we employ a variation of the DSatur algorithm~\cite{lewisBoundsConstructiveHeuristics2021} that maintains its polynomial time complexity. %
The DSatur algorithm is a greedy algorithm that iteratively selects the edge with the most different adjacent colors and selects the least admissible color for it.
The least admissible color is the smallest color that is different from all adjacent colors of the edge and larger than any color of an adjacent edge that does not share the same node in the independent set.
This ensures that the \cref{non-crossing constraint} is satisfied.
We modify the DSatur algorithm accordingly as defined in \cref{alg:dsatur}.
The following \cref{exp:color edges} illustrates its application.

\begin{algorithm}[h]
    \caption{Modified DSatur Algorithm}
    \label{alg:dsatur}
    \DontPrintSemicolon
    \SetCommentSty{textit}
    \SetNlSty{}{}{}
    \SetKwComment{tcp}{\(\triangleright\ \)}{}
    \KwIn{Interaction graph \(G = (V, E)\) (undirected)\newline
        Independent Set \(I\subseteq V\)}
    \KwOut{Valid coloring \(c:E\rightharpoonup \N\) of the edges}
    \For{\(e \in E\)}{\(e.\text{color}\gets\text{undef.}\)}
    \For{\(v\in I\) sorted by degree in descending order}{
        \(S\gets\) set of edges adjacent to \(v\)\; %
        \For{\(e\in S\) lexicographically ordered by the number of different adjacent colors (desc.) and degree (desc.)}{
            \(e.\text{color}\gets\) least admissible color\label{alg:line:least admissible color}\; \tcp*[r]{must be different from all adjacent colors and larger than any color of an adjacent edge not adjacent to \(v\)}
            \(S\gets S\setminus\{e\}\)\;
        }
    }
    \Return{\(c:e\mapsto e.\text{color}\)}
\end{algorithm}

\begin{figure}[tb]
    \centering
    \includegraphics[width=\linewidth]{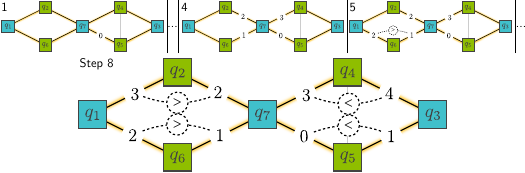}
    \caption{
        The coloring requires eight iterations, and the result is shown in the bottom row.
        The top row shows selected intermediate steps.
        The \enquote{\(<\)}-signs denote the extra constraints on the coloring to be valid.
    }
    \label{fig:color edges}
\end{figure}

\begin{figure*}[t]
    \centering
    \begin{subfigure}[t]{189pt}
        \centering
        \includegraphics[width=189pt]{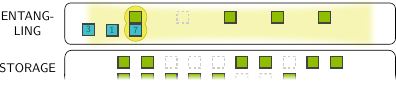}
        \caption{\(t=0\)}
    \end{subfigure}
    \hfill
    \begin{subfigure}[t]{160pt}
        \centering
        \includegraphics[width=160pt]{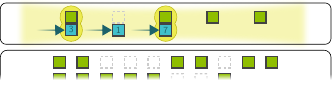}
        \caption{\(t=1\)}
    \end{subfigure}
    \hfill
    \begin{subfigure}[t]{160pt}
        \centering
        \includegraphics[width=160pt]{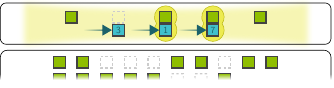}
        \caption{\(t=2\)}
    \end{subfigure}
    \\
    \begin{subfigure}[t]{189pt}
        \centering
        \includegraphics[width=189pt]{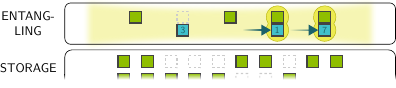}
        \caption{\(t=3\)}
    \end{subfigure}
    \hfill
    \begin{subfigure}[t]{160pt}
        \centering
        \includegraphics[width=160pt]{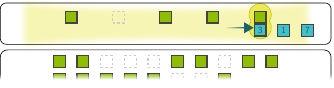}
        \caption{\(t=4\)}
    \end{subfigure}
    \hfill
    \begin{subfigure}[t]{160pt}
        \centering
        \includegraphics[width=160pt]{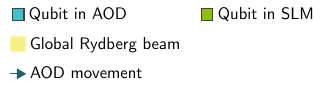}
    \end{subfigure}
    \caption{
        Visualization of all time steps corresponding to the run stemming from the interaction graph in \cref{exp:interaction graph}.
    }
    \label{fig:sequence}
\end{figure*}

\begin{example}[Coloring Edges]\label{exp:color edges}
    \Cref{alg:dsatur} starts with the node of the independent set with the highest degree and colors all its adjacent edges.
    Since, at this stage, no other edges are colored, they are colored with 1, 2, 3, and 4, respectively (see steps \(t=1\) and \(4\) in \cref{fig:color edges}).
    The two remaining nodes in the independent set have the same degree; we assume that it is the left one's turn next.
    For the edge colored in Step 5, the algorithm selects Color 2, the least color larger than the adjacent 1.
    The algorithm continues coloring the remaining edges, which results in the coloring shown in the bottom row in \cref{fig:color edges}.
\end{example}

With the coloring at hand, we can now calculate the exact positions of each qubit in every step and their movements in between.

\begin{figure}[tb]
    \centering
    \includegraphics[width=\linewidth]{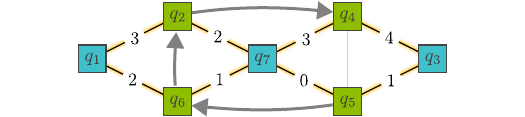}
    \caption{
        The grey arrows depict a directed graph that represents a partial order on the set of SLM qubits.
        The order of the qubits in which they are placed in the entangling zone, from left to right, is given by a topological ordering of the graph.
    }
    \label{fig:topological order}
\end{figure}

\subsection{Positioning the Qubits}\label{subsec:extracting solution}
Recall that the independent set determines which qubits are held in AOD traps, and the coloring determines the steps at which an entangling gate is executed.
We still need to determine the exact positions of the qubits in the entangling zone for each step to facilitate the correct entangling operation at the right time.
In the following, we will refer to qubits held in the AOD as AOD qubits and to those kept in SLM traps as SLM qubits.

The AOD qubits will move from left to right during one run.
Hence, the SLM qubit adjacent to one AOD qubit must be placed in ascending order with respect to the color assigned to the edge from the AOD to the SLM qubit.
This defines a total order on the SLM qubits in the neighborhood of one AOD qubit.
Combining the total orders induced by all AOD qubits gives a partial order on the set of all SLM qubits.%
\footnote{
    It is not self-evident that the directed graph induced by combining all total orders around an SLM qubit is acyclic, which is a requirement for the partial order to be well-defined.
    This limitation must already be taken care of during coloring: The inner for-loop actually orders the edges first by the partial order induced by the already colored edges and then by the number of different adjacent colors and their degree.
    Then, the function \texttt{leastAdmissibleColor} ensures that the new color does not introduce any cycles.
    \label{ftn:partial order}
}
We compute a total order on the SLM qubits by topologically sorting the directed graph induced by the partial order.
Thereby, the computed total order satisfies the partial order in the sense that the relation is a superset of the partial order.

\begin{example}[Topological Order]\label{exp:topological order}
    The colors around \(q_7\) in \cref{fig:topological order} induce an order on the SLM qubits.
    The order is depicted with grey arrows.
    The orderings induced by \(q_1\) and \(q_3\) are already subsumed by the existing order (also see \cref{ftn:partial order}).
    Following the topological order of the SLM qubits, we place the qubits in the entangling zone from left to right.
    Consequently, we place from left to right the qubits \(q_5\), \(q_6\), \(q_2\), and \(q_4\).
    This way, \(q_7\) can first interact with \(q_5\) in time step \(t=0\) and afterward with the other SLM qubits by moving one position right between the steps.
\end{example}

The order of the adjustable qubits in the AOD must be the same as the one used for the coloring, whereby the AOD qubit with the highest degree is placed rightmost.
By keeping an equal and sufficient distance between the SLM qubits to avoid any undesired interaction, we could, in principle, calculate the exact positions of all qubits in the entangling zone for each step.
However, there are time steps in which some AOD qubits shall not interact with any SLM qubit.
In this case, we need to ensure with so-called \emph{resting positions} that these AOD qubits are outside any other qubit's interaction radius.

\begin{example}[Resting Positions]\label{exp:resting positions}
    According to the coloring in \cref{fig:color edges}, the qubit \(q_1\) does not interact with any other qubit during time step \(t=1\).
    However, the order of the AOD qubits used for the coloring implies that \(q_1\) is located in between \(q_7\) and \(q_3\).
    In time step \(t=1\) those interact with \(q_5\) and \(q_6\), respectively.
    To avoid any interaction, we need to create a resting position for \(q_1\) between \(q_5\) and \(q_6\).
    Those resting positions can be reused later for other qubits.
    \begin{figure}[H]
        \centering
        \vspace*{-8pt}
        \includegraphics[width=\linewidth]{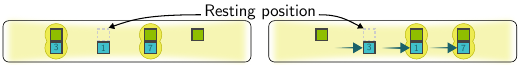}\\
        \vspace*{-6pt}
    \end{figure}
\end{example}

These resting positions shift all subsequential SLM qubits by one position to the right.
After this, the positions of the SLM qubits will be known.
Relative to those, the AOD qubits' positions in each time step are calculated according to the coloring.

\begin{example}[Complete Run]\label{exp:sequence}
    As seen in previous examples, in particular, \cref{exp:interaction graph}, \(q_7\) interacts with four qubits.
    Those are not part of the independent set and, hence, are placed in SLM traps.
    From one time step to the next, the AOD qubits are moved from left to right to match the position of their interacting partner.
    In time step \(t=1\), \(q_1\) stays at the resting position between qubit \(q_5\) and \(q_6\).
    This resting position will be reused for qubit \(q_3\) in the next step.
    All resulting steps of this run are shown in \cref{fig:sequence}.
\end{example}

\begin{figure*}[tb]
    \centering
    \includegraphics[width=\linewidth]{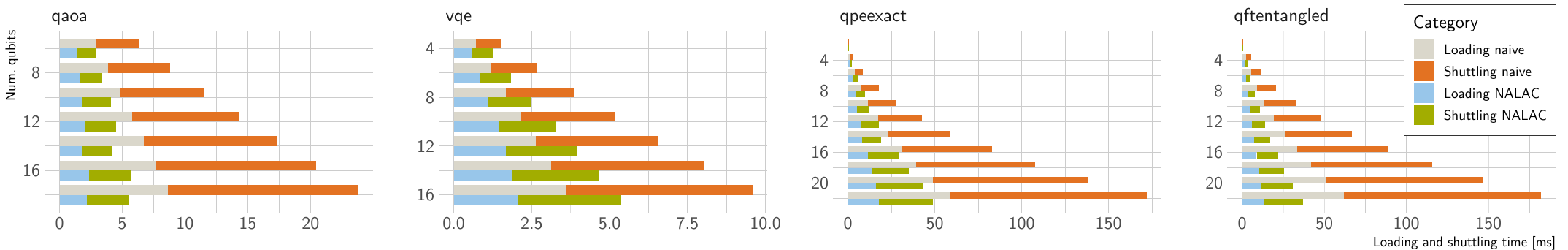}
    \caption{Comparison of the execution time of the circuit of the naive approach and NALAC.}
    \label{fig:execution time}
\end{figure*}

\subsection{Finalizing the Solution}\label{subsubsec:shuttling}
We are now able to perform one run of multiple entangling operations without additional load and store operations.
The only step missing is moving the affected qubits from the storage zone to the entangling zone and moving them back after the run.

To move the qubits to the entangling zone and store them in the previously computed order, we opt for a procedure that tries to maximize the number of qubits that can be loaded in parallel.
Unlike the naive solution, the proposed solution does not place the qubits from the start.
Instead, it follows a demand-driven approach: When a qubit is loaded into an AOD, its position can either be defined due to a previous run or be undefined.
In the latter case, we find a free spot in the storage zone to maximize the number of qubits that can be loaded in parallel.
Qubits can be loaded in parallel if they are in the same row and in the correct order, \ie, the order in which they are needed in the entangling zone.
Hence, the proposed solution finds the next free spot in the storage zone for a qubit with an undefined position satisfying both conditions.\footnote{We only execute circuits with as many qubits as the storage zone can fit. Hence, there will always be a free spot.}
To keep the space consumed horizontally outside of the SLM traps in the storage zone small, we first load qubits into the AOD with a large misplacement, meaning they are located far left and must travel far right or vice versa.
After a run has finished, the used qubits are moved back to the storage zone.
For this, we look for a minimal number of rows in the storage that fit all qubits, and the free spots in those rows are filled with qubits from the entangling zone.

Finally, as described in \cref{sec:ftqc}, the solution for the logical qubit is translated to a solution for arrays of atoms, each representing a logical qubit.
Therefore, every qubit in the logical solution is replaced by its respective array of atoms, where the logical qubit determines the upper left atom in the array.

\section{Evaluation}\label{sec:evaluation}
In order to evaluate the effectiveness of the proposed solution compared to the naive approach, both techniques have been implemented on top of the tool \emph{MQT QMAP} publicly available at \url{https://github.com/cda-tum/mqt-qmap}.
The resulting tool is called NALAC, short for Neutral Atom Logical Array Compiler.

\begin{wrapfigure}{R}{\dimexpr.5\linewidth-.5\columnsep\relax}
    \centering
    \includegraphics[width=\linewidth]{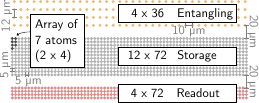}
    \caption{
        Layout of the zoned architecture used for the experiments. %
    }
    \label{fig:architecture wide}
\end{wrapfigure}
As a basis for our experiments, we used the architecture as illustrated in \cref{fig:architecture wide}.
To compute the different routing overheads produced by both approaches, we used the following parameters based on the specifications of the neutral atom architecture in~\cite{bluvsteinLogicalQuantumProcessor2023}:
\begin{table}[H]
    \vspace{-4pt}
    \renewcommand{\arraystretch}{1.3}
    \begin{tabularx}{\linewidth}{XXX}
        \toprule
        \textbf{Load/Store} & Duration [\(\SI{}{\micro\s}\)]           & \(20\)   \\
        \textbf{Shuttling}  & Speed [\(\SI{}{\micro\m\per\micro\s} \)] & \(0.55\) \\
        \textbf{CZ Gate}    & Duration [\(\SI{}{\micro\s}\)]           & \(0.2\)  \\
        \bottomrule
    \end{tabularx}
    \vspace{-4pt}
\end{table}

As input circuits, we use preprocessed versions of circuits taken from the MQT Bench library~\cite{quetschlichMQTBenchBenchmarking2022} grouped into circuit families, such as qaoa, vqe, qft, etc.
The preprocessing consists of translating the original circuits to the gate set supported by the neutral atom architecture, \ie, local as well as global rotations, and CZ gates, using Qiskit~\cite{Qiskit} and a custom Python script.
All experiments were conducted on a machine with an Apple M3 chip and 16 GB of RAM.

\subsection{Comparison of the Naive Solution to NALAC}\label{subsec:comparison}

\begin{table}[b]
    \centering
    \caption{Naive Solution vs. NALAC}
    \label{tab:other execution times}
    \def\centerCol#1{\multicolumn{1}{c}{#1}} %
    \def\idleCol{\multicolumn{2}{c}{\textbf{Idle Rydberg Beams}}} %
    \begin{tabularx}{\linewidth}{Xrrrr}
        \toprule
        \centerCol{\textbf{Circuit}}                  & \centerCol{\textbf{Num.}}     & \multicolumn{2}{c}{\textbf{Routing}}                       & \centerCol{\(\varnothing\) \textbf{Parallel}}                                \\
        \centerCol{\textbf{(20 Qubits)}}              & \centerCol{\textbf{CZ-Gates}} & \multicolumn{2}{c}{\textbf{Overhead [\(\SI{}{\milli\s}\)]}} & \centerCol{\textbf{CZ-Gates}}                                                \\
                                                      &                               & \centerCol{\textbf{Naive}}                                 & \centerCol{\textbf{NALAC}}                    & \centerCol{\textbf{(NALAC)}} \\
        \midrule
        ae                                            & 380                           & 130                                                        & 32                                          & 1.1                          \\
        dj                                            & 19                            & 7                                                       & 1                                           & 1.0                          \\
        ghz                                           & 19                            & 6                                                       & 7                                           & 1.0                          \\
        graphstate                                    & 20                            & 7                                                       & 1                                           & 3.3                          \\
        portfoliovqe                                  & 570                           & 195                                                        & 31                                          & 1.0                          \\
        qft                                           & 408                           & 139                                                        & 22                                          & 1.0                          \\
        qftentangled                                  & 429                           & 147                                                        & 31                                          & 2.8                          \\
        qnn                                           & 778                           & 266                                                        & 57                                          & 3.6                          \\
        qpeexact                                      & 406                           & 139                                                        & 43                                          & 3.7                          \\
        qpeinexact                                    & 407                           & 139                                                        & 43                                          & 3.7                          \\
        \multicolumn{2}{l}{realamprandom \hfill 570}  & 195                           & 27                                                       & 1.2                                                                          \\
        su2random                                     & 570                           & 195                                                        & 27                                          & 1.2                          \\
        \multicolumn{2}{l}{twolocalrandom \hfill 570} & 195                           & 27                                                         & 1.2                                                                          \\
        wstate                                        & 38                            & 13                                                         & 7                                           & 1.0                          \\
        \bottomrule
    \end{tabularx}
\end{table}

To demonstrate the advantage of NALAC in terms of the two objectives discussed in \cref{sec:proposed solution}, namely
\begin{enumerate*}
    \item improved parallel gate execution, and
    \item improved parallel AOD shuttling,
\end{enumerate*}
we evaluated the time taken for loading, shuttling, and storing when executing the resulting circuits---short the \emph{routing time overhead}.
\Cref{fig:execution time} shows the obtained loading (including storing) and shuttling times separately for four different circuit families in various sizes.
As one can see, NALAC produces circuits with significantly shorter loading and shuttling times compared to the naive solution.
\Cref{tab:other execution times} summarizes the results for those four and additional circuit families with a fixed number of 20 logical qubits.
The last column shows the average number of entangling operations performed in parallel per Rydberg beam for circuits compiled with NALAC.
Note that the corresponding number for the naive solution is always one.
The low numbers for the average parallel CZ gates for some circuits are not surprising, as the CZ gates in these circuits are interleaved with one-qubit gates, which prevent the parallel execution of many CZ gates.
In general, it can be said that the more parallelism the structure of a circuit allows, the more NALAC can exploit its advantage over the serial naive approach.
In the ghz circuit, for example, NALAC even has a non-significant larger routing overhead because---due to the structure of the circuit---the execution of the CZ gates is interrupted by one-qubit gates, which forces the CZ gates to be executed serially.
Overall, those figures demonstrate that NALAC is superior to the naive solution because it yields substantially less routing time overhead and higher gate parallelism.

\subsection{Comparison of Compilation Times}\label{subsec:routing time}
\begin{wrapfigure}{R}{\dimexpr.5\linewidth-.5\columnsep\relax}
    \centering
    \includegraphics[width=\linewidth]{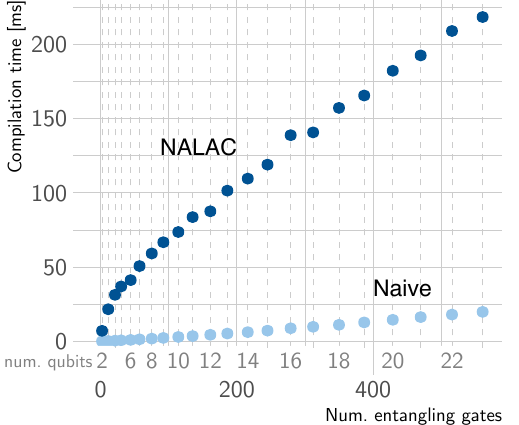}
    \caption{Compilation time of the naive approach and NALAC.}
    \label{fig:compilation time}
\end{wrapfigure}
The improved execution times of NALAC come at a price: The proposed approach is more complex than the naive one and requires a longer compilation time.
To evaluate the cost of the improvement, we compared the compilation time of both approaches.
\Cref{fig:compilation time} shows the compilation time for the naive solution and NALAC for circuits of the \texttt{qftentangled} family with different numbers of qubits.
As expected, the runtime of NALAC is significantly higher than that of the naive solution.
However, the performance remains polynomial; even for larger circuits, the runtime of NALAC is still within the range of milliseconds, making it a suitable approach even for larger numbers of qubits.

\subsection{Influence of Array Size}\label{subsec:array size}
\begin{figure}[b]
    \centering
    \begin{subfigure}[t]{124pt}
        \centering
        \includegraphics[width=\linewidth]{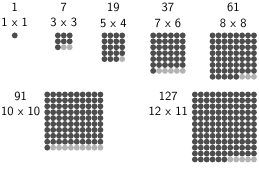}
        \caption{The size of the arrays for different color codes.}
        \label{fig:code sizes}
    \end{subfigure}
    \hfill
    \begin{subfigure}[t]{124pt}
        \centering
        \includegraphics[width=\linewidth]{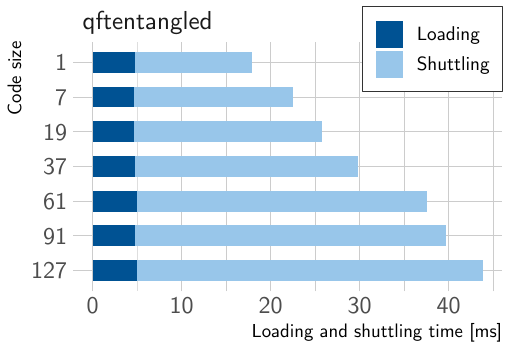}
        \caption{
            Impact of increasing array sizes on the routing overhead.
        }
        \label{fig:array size}
    \end{subfigure}
    \caption{Comparing different array sizes.}
\end{figure}

The experiments above were all performed with a logical array size of one, \ie, one atom represents one logical qubit.
As argued in \cref{sec:ftqc}, NALAC also supports larger atom arrays representing a logical qubit.
To evaluate that, also the impact of the array size on the routing overhead is considered, shown in \cref{fig:array size}.
As can be seen, the qubits need to move further with larger array sizes, increasing the shuttling times.
However, the loading times remain roughly the same since all qubits in one array can be loaded in parallel.
The only effect that can happen here is that fewer logical qubits fit in one row, which requires more loading steps and consequently slightly increases the loading time.

\subsection{Supporting Hardware Design}\label{subsec:hardware design}
\begin{figure}[t]
    \centering
    \begin{subfigure}[t]{124pt}
        \centering
        \includegraphics[width=\linewidth]{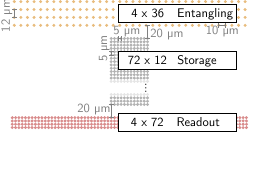}
        \caption{
            Architecture with a narrow storage zone.
        }
        \label{fig:architecture narrow}
    \end{subfigure}\hfill
    \begin{subfigure}[t]{124pt}
        \centering
        \includegraphics[width=\linewidth]{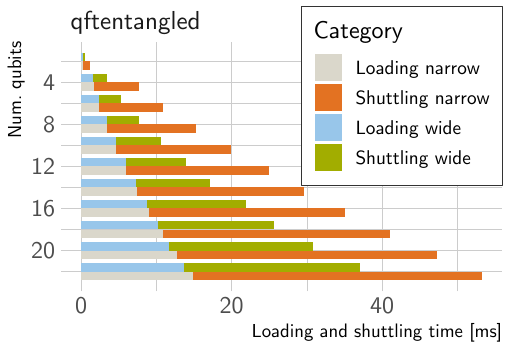}
        \caption{Impact of a wide and a narrow storage zone on the routing overhead.}
        \label{fig:narrow wide}
    \end{subfigure}
    \caption{Comparing different shapes of the storage zone.}
\end{figure}

Finally, in a last series of experiments, we investigated how the proposed approach can even be used to evaluate different hardware designs.
To this end, we exemplarily considered another architecture, namely the one shown in \cref{fig:architecture narrow}, which has a narrower storage zone than the one from \cref{fig:architecture wide} considered before.
\Cref{fig:narrow wide} shows the corresponding resulting routing time overhead---clearly showcasing the impact of this design.
In fact, the results show that the narrow storage zone leads to both longer loading and shuttling times.
This can be explained by the fact that
\begin{enumerate*}[label=(\arabic*)]
    \item the qubits need to travel longer distances on average to reach the entangling zone, and
    \item fewer qubits fit in one row in the storage zone.
\end{enumerate*}
This demonstrates how the proposed approach can help hardware designers to analyze trade-offs between different hardware parameters and settings.

\section{Conclusions}\label{sec:conclusions}
In this work, we considered the development of target-specific quantum compilers specialized for zoned neutral atom architectures.
We first provided an abstract model of zoned neutral atom architectures, which are a promising candidate for fault-tolerant gate execution and, therefore, general FTQC.
Then, we proposed a novel solution to the routing problem of entangling operations to this architecture.
This solution minimizes the time required for loading, shuttling, and storing the qubits while maximizing the gate parallelism of entangling gates.
We implemented the proposed solution in the tool NALAC and compared it to a naive solution.
Evaluations showed that NALAC efficiently routes entangling operations of even larger quantum circuits to the zoned neutral atom architecture, applies to various array sizes, and can even support designers evaluating different hardware designs.
NALAC is publicly available in open-source as part of the \emph{Munich Quantum Toolkit}~(MQT,~\cite{mqt}) at \mbox{\url{https://github.com/cda-tum/mqt-qmap}}.
For full FTQC, this tool needs to be extended with further features, like magic state injection~\cite{fowlerSurfaceCodesPractical2012} and error correction cycles~\mbox{\cite{shorFaulttolerantQuantumComputation1996,gottesmanStabilizerCodesQuantum1997}}.

\subsection*{Acknowledgements}\label{sec:ack}
We want to thank Johannes Zeiher for providing us with helpful comments on a draft of this article and the anonymous reviewers for their helpful feedback.

This work received funding from the European Research Council (ERC) under the European Union’s Horizon 2020 research and innovation program (grant agreement No. 101001318), was part of the Munich Quantum Valley, which the Bavarian state government supports with funds from the Hightech Agenda Bayern Plus, and has been supported by the BMWK based on a decision by the German Bundestag through project QuaST, as well as by the BMK, BMDW, and the State of Upper Austria in the frame of the COMET program (managed by the FFG).

\printbibliography

\end{document}
\typeout{get arXiv to do 4 passes: Label(s) may have changed. Rerun}